\documentclass[onecolumn]{revtex4}

\usepackage{graphicx}
\usepackage{dcolumn}
\usepackage{bm}

\input epsf

\begin{document}

\title{\Large A Dark Energy Model with Generalized Uncertainty Principle in the Emergent,
\emph{Intermediate} and \emph{Logamediate} Scenarios of the
Universe}

\author{\bf Rahul Ghosh$^1$\footnote{ghoshrahul3@gmail.com}, Surajit
Chattopadhyay$^2$\footnote{surajit$_{_{-}}2008$@yahoo.co.in,
surajit.chattopadhyay@pcmt-india.net} and Ujjal
Debnath$^3$\footnote{ujjaldebnath@yahoo.com,
ujjal@iucaa.ernet.in}}

\affiliation{$^1$ Department of Mathematics, Bhairab Ganguly
College, Kolkata-700 056, India.\\
$^2$Department of Computer Application (Mathematics Section),
Pailan College of Management and Technology, Bengal Pailan Park,
Kolkata-700 104, India.\\
$^3$Department of Mathematics, Bengal Engineering and Science
University, Shibpur, Howrah-711 103, India.}

\date{\today}

\begin{abstract}
This work is motivated by the work of Kim et al (2008), which
considered the equation of state parameter for the new agegraphic
dark energy based on generalized uncertainty principle coexisting
with dark matter without interaction. In this work, we have
considered the same dark energy interacting with dark matter in
emergent, \emph{intermediate} and \emph{logamediate} scenarios of
the universe. Also, we have investigated the statefinder, kerk and
lerk parameters in all three scenarios under this interaction. The
energy density and pressure for the new agegraphic dark energy
based on generalized uncertainty principle have been calculated
and their behaviors have been investigated. The evolution of the
equation of state parameter has been analyzed in the interacting
and non-interacting situations in all the three scenarios. The
graphical analysis shows that the dark energy behaves like
quintessence era for logamediate expansion and phantom era for
emergent and intermediate expansions of the universe.
\end{abstract}


\maketitle

\section{\normalsize\bf{Introduction}}

Cosmological observations suggest that our universe is currently
undergoing a phase of accelerated expansion driven by some unknown
energy component characterized by negative pressure (Perlmutter,
et al., 1999;  Bachall et al, 1999; Copeland et al, 2006).
Recently, the combination of WMAP3 and Supernova Legacy Survey
data shows a significant constraint on the equation of state (EOS)
for the dark energy, $w_{ob}=-0.97^{+0.07}_{-0.09}$ in a flat
universe (Seljak et al, 2006). This unknown energy component is
dubbed as ``dark energy" and a great variety of models have been
proposed so far to describe this dark energy.  Observations show
that the energy density of DE occupies about $70\%$ of today's
universe (Cai et al, 2010). However, at early cosmological epochs
DE could not have dominated since it would have destroyed the
formation of the observed large scale structure. These features
have significantly challenged our thoughts about Nature. People
begin to ask questions like (Cai et al, 2010): What is the
constitution of DE? Why it dominates the evolution of our universe
today? What is the relation among DE, dark matter and particle
physics, which is successfully constructed?  Some recent reviews
on dark energy are Copeland et al (2006), Padmanabhan (2005,
2006), Sahni and Starobinsky (2006) and Sahani (2005).  The
simplest candidate of dark energy is a tiny positive cosmological
constant. However, as is well known, it is plagued by the
so-called ``cosmological constant problem" and ``coincidence
problem" (Copeland et al, 2006). Other dark energy models include
quintessence (Ratra and Peebles, 1988), phantom (Nojiri et al,
2005), quintom (Guo et al, 2005; Elizalde et al, 2004), Chaplygin
gas (Gorini et al, 2003), tachyon (Calcagni and Liddle, 2006),
hessence (Wei et al, 2005), Ricci dark energy (Feng, 2008), and
electro magnetic dark energy (Beck et al, 2008). There are two
other candidates of dark energy based on holographic principle
(Bousso, 2002). They are holographic dark energy (Li, 2004) and
agegraphic dark energy model (Wei and Cai, 2008). The first is
based on the Bekenstein-Hawking energy bound $E_{\Lambda}\leq
E_{BH}$ with the energy $E_{BH}$ of a universe-sized black hole
which produces $L^{3}\rho_{\Lambda}\leq m_{p}^{2}L$ with the
length scale $L$ (IR cutoff) of the universe and the Planck mass
$m_{p}$. The largest $L$ allowed is the one saturating this
inequality, thus the holographic dark energy density is
$\rho_{\Lambda}=3c^{2}m_{p}^{2}L^{-2}$(Li, 2004; Pavon and
Zimdahl, 2005), where $c^{2}$ is a constant. The later is based on
the Karolyhazy relation of $\delta t$ and the time-energy
uncertainty of $\Delta E\sim t^{-1}$ in the Minkowiski spacetime
with a given time scale $t$, which gives $\rho_{q}\sim\frac{\Delta
E}{(\delta t)^{3}}\sim\frac{m_{p}^{2}}{t^{2}}$ (Maziashvili,
2007). Nojiri and Odintsov (2006) suggested generalized
holographic dark energy where infrared cutoff is identified with
combination of FRW parameters: Hubble constant, particle and
future horizons, cosmological constant and universe life-time (if
finite). This study of Nojiri and Odintsov (2006) also reviewed
other known holographic dark energy models.
\\\\
The problem of discriminating different dark energy models is now
emergent. In order to solve this problem, a sensitive and robust
diagnostic for dark energy is a must. The statefinder parameter
pair ${r, s}$ introduced by Sahni et al (2003) and Alam et al
(2003) is proven to be useful tools for this purpose. The
statefinder pair is a `geometrical' diagnostic in the sense that
it is constructed from a space-time metric directly, and it is
more universal than `physical' variables which depends upon
properties of physical fields describing dark energy, because
physical variables are, of course, model-dependent (Feng, 2008).
Details of the statefinder parameters would be discussed in the
subsequent section. The spatially flat $\Lambda$CDM scenario
corresponds to a fixed point $\{r,s\}=\{1,0\}$ in  the $r$-$s$
plane. The statefinder can successfully differentiate between a
wide variety of dark energy models including the cosmological
constant, quintessence, the Chaplygin gas, braneworld models and
interacting dark energy models. The statefinder diagnostics have
been investigated for tachyonic field (Chattopadhyay et al, 2008),
holographic dark energy (Zhang, 2005), Ricci dark energy (Feng,
2008), quintessence (Zhang, 2005), Yang-Mills dark energy (2008),
quintom dark energy (Wu and Yu, 2005), dilaton dark energy (Huang
et al, 2008). In a study on interacting new agegraphic dark energy
by Zhang et al (2010) it was found that the $r$-$s$ trajectory is
confined in the first quadrant of the $r$-$s$ plane for various
forms of interaction. Statefinders generalize such well known
observational characteristics of the expansion as the Hubble
(first-order) and the deceleration (second-order) parameters. The
expansion factor or scale factor $a$ of the universe can be Taylor
expanded around the present epoch $t_{0}$ as
$a(t)=a_{0}\left[1+\sum_{n=1}^{\infty}\frac{A_{n}(t_{0})}{n!}\{H_{0}(t-t_{0})\}^{n}\right]$;
where $A_{n}=\frac{a^{(n)}}{H^{n}},~~n\in N$ (Arabsalmania and
Sahni, 2011). Here, $a^{(n)}$ is the $n$-th derivative of the
scale factor with respect to time. For various values of $n$, we
get different parameters like  jerk `$j$' , snap `$s$', lerk
`$l$', etc (Visser, 2005; Arabsalmania and Sahni, 2011; Dabrowski,
2005). It should be mentioned that for $n=3$ we get statefinder
parameter `$r$', which is also known as `jerk' parameter
(Arabsalmania and Sahni, 2011).
\\\\

In this work, we consider the new agegraphic dark energy model
with the generalized uncertainty principle (GUP). This work is
motivated by the work of Kim et al (2008), who were first consider
the new agegraphic dark energy models with the GUP. The GUP and
its consequences has been discussed in the papers like Garay
(1995), Scardigli (1999) and Rama (2001). Although the GUP has its
origins in the string theory, may play a role of evolution of the
universe Kim et al (2008). In this paper we consider the
interacting new agegraphic dark energy model with the generalized
uncertainty principle in emergent, \emph{intermediate} and
\emph{logamediate} scenarios of the universe. We investigate the
behavior of the equation of state parameter, statefinder, kerk and
lerk parameters under this interaction and also the fate of the
universe through statefinder diagnostics.
\\\\

\section{New agegraphic dark energy model with generalized uncertainty principle (GUP)}

The metric of a spatially flat homogeneous and isotropic universe
in FRW model is given by

\begin{equation}
ds^{2}=dt^{2}-a^{2}(t)\left[dr^{2}+r^{2}(d\theta^{2}+sin^{2}\theta
d\phi^{2})\right]
\end{equation}

where $a(t)$ is the scale factor. The Einstein field equations are
given by

\begin{equation}
H^{2}=\frac{1}{3}\rho
\end{equation}
and
\begin{equation}
\dot{H}=-\frac{1}{2}(\rho+p)
\end{equation}

where $\rho$ and $p$ are energy density and isotropic pressure
respectively (choosing $8\pi G=c=1$).\\

The conservation equation is given by

\begin{equation}
\dot{\rho}+3H(\rho+p)=0
\end{equation}

Next, we consider the interaction between the new agegraphic dark
energy using GUP and dark matter. According to the GUP, the energy
density is defined by (Kim et al, 2008)

\begin{equation}
\rho_{G}=\frac{\Delta E_{G}}{(\delta t)^{3}}
\end{equation}

where, where $\delta t$ is given by the
K$\acute{a}$rolyh$\acute{a}$zy relation of time fluctuations as
$\delta t=t_{p}^{2/3}t^{1/3}$. Solving the saturation of the GUP
leads to

\begin{equation}
\Delta E_{G}=\frac{1}{t}+\frac{\zeta}{t^{3}}
\end{equation}

labelling of $\zeta$ and $t_{p}$ are
\begin{equation}
\zeta=\left(\frac{\xi}{n}\right)^{2}~~,~~~~t_{p}^{2}=\frac{1}{3n^{2}m_{p}^{2}}
\end{equation}
 Consequently, the dark energy density is described with two
parameters $(n, \xi)$ as (Kim et al, 2008)

\begin{equation}
\rho_{G}=\frac{3n^{2}m_{p}^{2}}{t^{2}}+\frac{3\xi^{2}}{t^{4}}
\end{equation}

Wei and Cai (2008) proposed the new agegraphic dark energy model
characterized by the energy density

\begin{equation}
\rho_{A}=\frac{3n^{2}m_{p}^{2}}{\eta}
\end{equation}

where the conformal time $\eta$ is given by

\begin{equation}
\eta=\int\frac{dt}{a}
\end{equation}

Using (10) in (8) the dark energy density based on GUP takes the
form (Kim et al, 2008)

\begin{equation}
\rho_{G}=\frac{3n^{2}m_{p}^{2}}{\eta^{2}}+\frac{3\xi^{2}}{\eta^{4}}
\end{equation}

Here, now we are considering interaction between dark matter and
the dark energy. That is why, in equations (2), (3) and (4) we
replace $\rho$ and $p$ by $\rho_{total}$ and $p_{total}$
respectively with

\begin{equation}
\rho_{total}=\rho_{G}+\rho_{m}~~,~~~~p_{total}=p_{G}+p_{m}
\end{equation}

where $p_{G}$, $p_{m}$ and $\rho_{m}$ denote the pressure of the
GUP based dark energy, pressure of dark matter and the density of
the dark matter respectively. Consequently, the conservation
equation (4) becomes

\begin{equation}
\dot{\rho}_{total}+3H(\rho_{total}+p_{total})=0
\end{equation}

As in the case of interaction the components do not satisfy the
conservation equation separately, we need to reconstruct the
conservation equation by introducing an interaction term $Q$. It
is important to note that the conservation equations imply that
the interaction term should be a function of a quantity with units
of inverse of time (a first and natural choice can be the Hubble
factor $H$) multiplied with the energy density. Therefore, the
interaction term could be in any of the forms (Sheykhi, 2010; Wei
and Cai, 2009): $Q\propto H\rho_{G}$, $Q\propto
H\rho_{m}$ and $Q\propto H\rho_{total}$.\\

Considering the interaction term $Q$ as $Q=3H\delta\rho_{m}$,
where $\delta$ is the interaction parameter, the conservation
equation (13) takes the form

\begin{equation}
\dot{\rho}_{G}+3H(\rho_{G}+p_{G})=Q
\end{equation}
and
\begin{equation}
\dot{\rho}_{m}+3H\rho_{m}(1+w_{m})=-Q
\end{equation}

where, $w_{m}=\frac{p_{m}}{\rho_{m}}$ is the equation of state
parameter for dark matter. It may be noted that similar choice of
the interaction term has been made in Wang et al (2005), Sheykhi
(2009). If $Q>0$, there is a flow of energy from dark matter to
dark energy (Cataldo et al, 2008). We are going to discuss the
said interactions in three scenarios:\\

\begin{enumerate}
    \item {\bf Emergent scenario} (Mukherjee et al, 2006), where the
    scale factor has the form $a(t)=a_{0}\left(B+e^{A t}\right)^{m}$
    with $a_{0}>0,~~A>0,~~B>0,~~m>1$.
    \item {\bf Intermediate scenario} (Barrow and Nunes, 2007; Barrow and Liddle,
    1993), where $a(t)=exp (\lambda t^{\beta})$ with
    $\lambda>0;~~0<\beta<1$.
    \item {\bf Logamediate scenario} (Barrow and Nunes, 2007), where $a(t)=exp(\mu(\ln
    t)^{\alpha})$ with $\mu\alpha>0,~~\alpha>1$.\\
\end{enumerate}

At this juncture it should be stated that some authors first
choose the scale factor in power law, exponential or in other
forms and then find out other variables with some conditions under
these solutions. This `reverse' way of investigations had earlier
been used extensively by Ellis and Madsen (1991) who chose various
forms of scale factor and then found out the other variables from
the field equations. Subsequently, this approach has been adopted
by Banerjee and Das (2005) who clearly stated ``This is not the
ideal way to find out the dynamics of the universe, as here the
dynamics is assumed and then the fields are found out without any
reference to the origin of the field. But in the absence of more
rigorous ways, this kind of investigations collectively might
finally indicate towards the path where one really has to search".
In another study, Feinstein (2002) assumed scale factor in the
power law form to model the potential by an inverse square law in
terms of the tachyon field. Campuzano et al (2010) studied the
curvaton reheating assuming assuming the scale factor in the
\emph{logamediate} scenario i.e. in the form given under item 3 of
the above list. So in particular, we have chosen the scale factor
in the forms enlisted above. Mukherjee et al (2006) obtained the
general solution of the scale factor for the emergent universe
without referring to the actual source of the energy density. So
we use the form under item $1$ of the above list as the choice of
scale factor for the emergent universe. Such choice of scale
factor has been used in the references like Mukherjee et al
(2005), Debnath (2008) and Paul and Ghose (2010). In the
particular scenario of `intermediate' inflation the expansion
scale factor of the Friedmann universe evolves as in item $2$; the
expansion of the Universe is slower for standard de Sitter, which
arises when $\beta=1$, but faster than in power-law inflation,
$a=t^{\beta}$, with $\beta>1$ constant (Barrow and Nunes, 2007).
This form of scale factor has been used in Khatua and Debnath
(2010). Barrow and Nunes (2007) considered `logamediate' inflation
where the cosmological scale factor expands in the form expressed
in the item 3 of the above list. This form of scale factor has
been used in Khatua and Debnath (2010). All of the above three
scenarios have been discussed at length in the references cited in
the above list. In this paper we are not going into the
descriptions of the said scenarios. For all of the said scenarios,
we have considered the characteristics of the universe expansion
as:

\begin{description}
    \item[The deceleration parameter:] The acceleration of
the universe can be quantified through a dimensionless
cosmological function known as the `deceleration parameter' $q$
given by (Dabrowski, 2005)
\begin{equation}
 q=-\frac{1}{H^{2}}\frac{\ddot{a}}{a}
 \end{equation}
where $q<0$ describes an accelerating universe, whereas $q\geq0$
for a universe which is either decelerating or expanding at the
`coasting' $a\propto t$ (Alam et al, 2003).
\\
    \item[The ``jerk" or ``statefinder" parameter:]
    It is known that ``jerk" parameter is another name of
    the ``statefinder" parameter. The statefinder parameter is given by
   (Alam et al, 2003; Arabsalmania and Sahni, 2011)
\begin{equation}
 r=\frac{1}{H^{3}}\frac{\dddot{a}}{a}
\end{equation}
     It is easy to see that $r$ is a natural next step beyond $H$ and $q$.
    We can easily see that this diagnostic is constructed from
the $a(t)$ and its derivatives up to the third order. So, the
statefinder probe is the expansion dynamics of the universe
through higher derivatives of the scale factor (Huang et al,
2008).
\\
    \item[The ``second statefinder" parameter:]
    The second statefinder parameter is given by (Alam et al, 2003)
\begin{equation}
    s=\frac{r-1}{3(q-\frac{1}{2})}
    \end{equation}
    By far, many models have been differentiated by this geometrical diagnostic method. Its
important property is that $\{r,s\}=\{1,0\}$ is a fixed point for
the flat $\Lambda$CDM FRW cosmological model (Huang et al, 2008).
Departure of a given DE model from this fixed point is a good way
of establishing the ``distance" of this model from flat
$\Lambda$CDM.\\

    \item[The ``kerk" or ``snap" parameter:] Snap, which involves the fourth
    time derivative of scale factor, is also sometimes called
    ``kerk". This parameter is given by
    \begin{equation}
    k=-\frac{1}{H^{4}}\frac{a^{(4)}}{a}
    \end{equation}
    This parameter has been discussed
    in the references like Dabrowski (2005), Dunajski and Gibbons (2008) and Arabsalmania and
    Sahni (2011) .
    \\
    \item[The ``lerk" parameter:] This parameter involves the fifth time derivative of scale
    factor. This parameter is given by (Dabrowski, 2005)
    \begin{equation}
    l=\frac{1}{H^{5}}\frac{a^{(5)}}{a}
    \end{equation}

\end{description}
In this work we have investigated all of the said parameters for
the new agegraphic dark energy based on generalized uncertainty
principle under the three different scenarios mentioned earlier.
\\

\section{Interaction in the emergent scenario}

In this section we consider the interaction between dark energy
and dark matter under emergent scenario and discuss the $r$-$s$
trajectories along with other parameters involving various orders
of derivative of the scale factor. For emergent scenario using
$a(t)=a_{0}\left(B+e^{A t}\right)^{m}$ in (10) we get
\begin{equation}
\eta=-\frac{\left(1+Be^{-At}\right)^{m}\left(B+e^{At}\right)^{-m}~_{2}F_{1}\left[m,m,1+m,-Be^{-At}\right]}{Aa_{0}m}
\end{equation}
Consequently, using (11) and (16) we get the dark energy density
as
\begin{equation}
\rho_{G}=\frac{3A^{4}a_{0}^{4}\left(1+Be^{-At}\right)^{-4m}\\
\left(B+e^{At}\right)^{4m}m^{4}\xi^{2}}{~_{2}F_{1}\left[m,m,1+m,-Be^{-At}\right]^{4}}
+\frac{3A^{2}a_{0}^{2}\left(1+Be^{-At}\right)^{-2m}\left(B+e^{At}\right)^{2m}n^{2}m^{2}m_{p}^{2}}{~_{2}F_{1}\left[m,m,1+m,-Be^{-At}\right]^{2}}
\end{equation}
From the conservation equation (15) we get under emergent scenario
\begin{equation}
\rho_{m}=\rho_{m0}\left[a_{0}\left(B+e^{At}\right)^{m}\right]^{-3(1+w_{m}+\delta)}
\end{equation}

\begin{figure}
\includegraphics[height=2.1in]{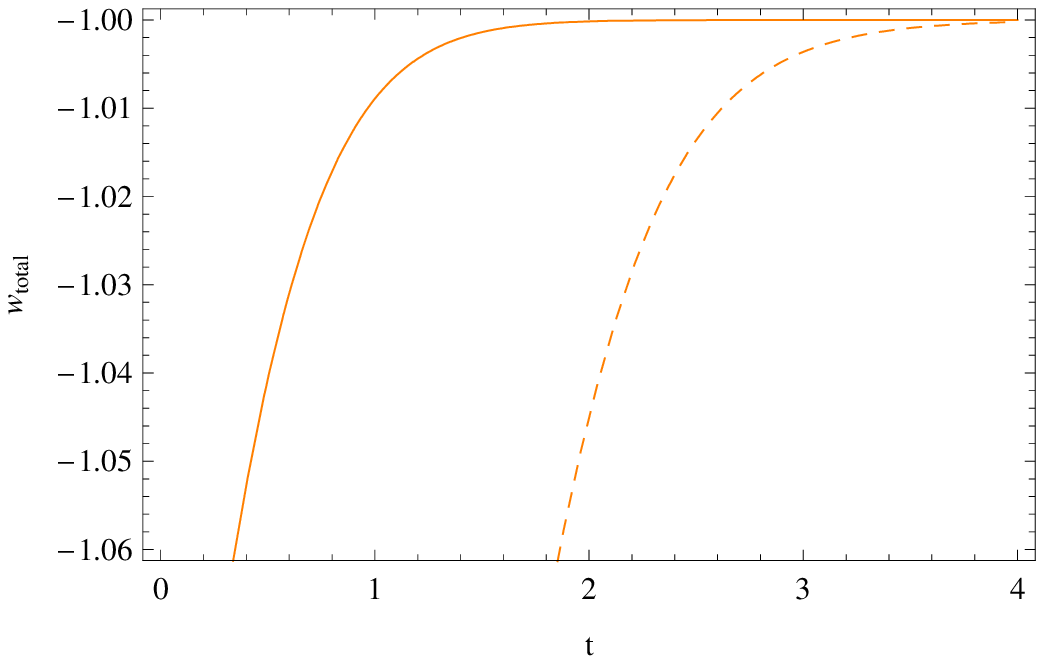}~~~~
\includegraphics[height=2.1in]{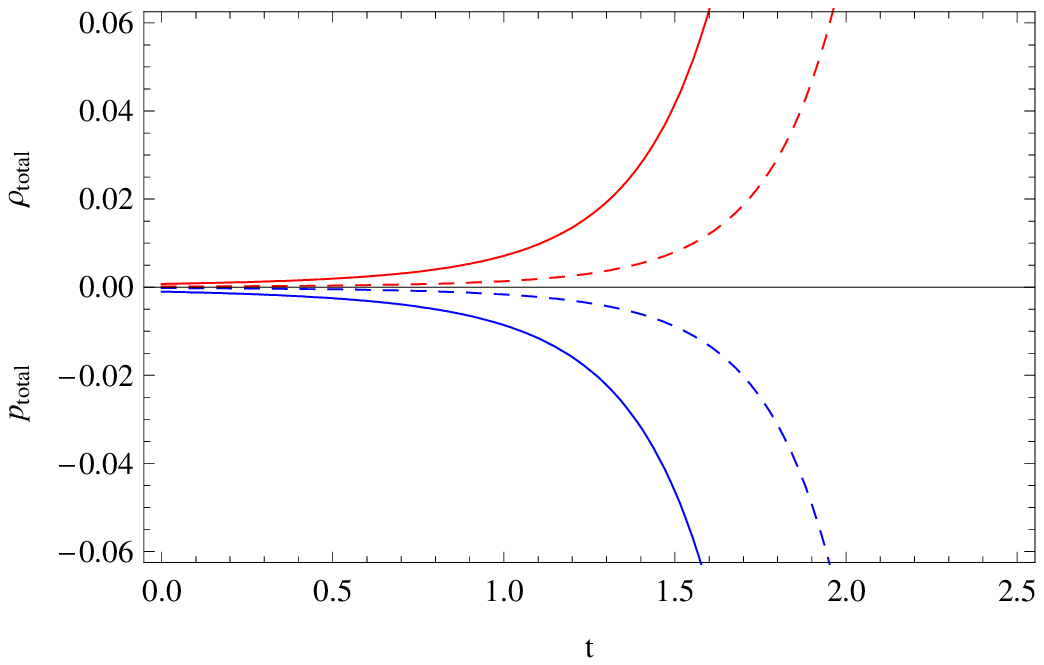}\\
\vspace{1mm} ~~~~~~~Fig.1~~~~~~~~~~~~~~~~~~~~~~~~~~~~~~~~~~~~~~~~~~~~~~~~~~~~~~~~~~~~~~~~~~~~~~~~~Fig.2~~~\\

\vspace{6mm} Fig. 1 shows the evolution of the equation of state
parameter $w_{total}$ in the interacting (thick line) and
non-interacting (dashed line) situations in the emergent universe
scenario.\\
 Fig. 2 shows the total energy density $\rho_{total}$ and pressure $p_{total}$ in the interacting
 (thick line) and non-interacting (dashed line) situations in the emergent universe scenario.\\

 \vspace{3mm}

 \end{figure}

 \begin{figure}
\includegraphics[height=2.1in]{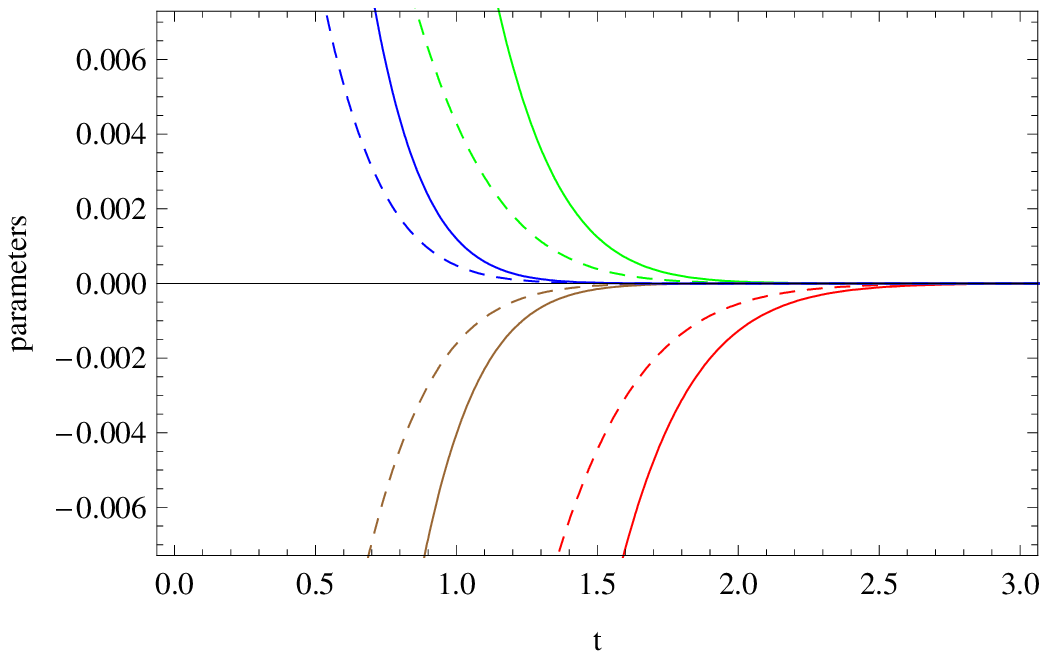}~~
\includegraphics[height=2.3in]{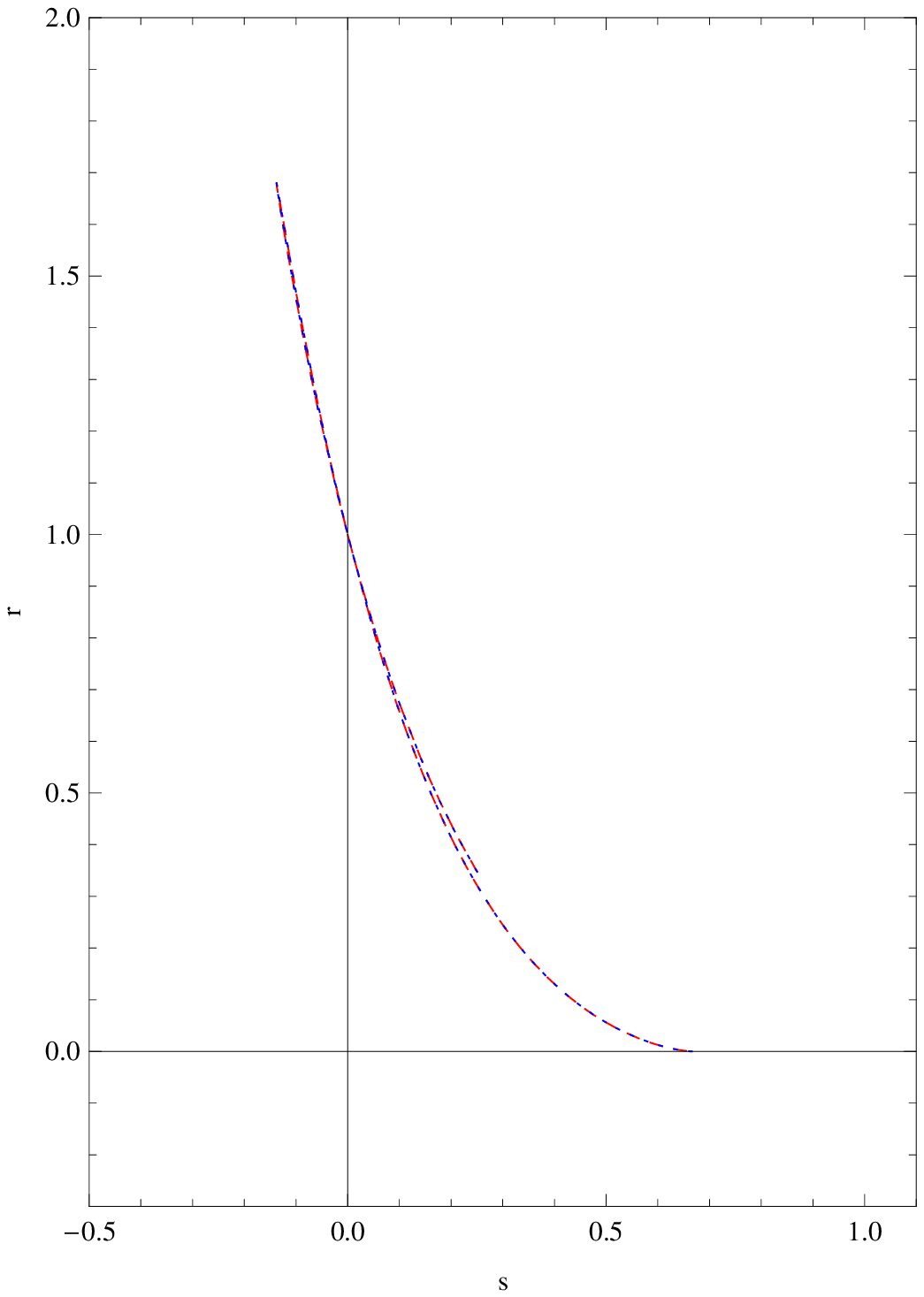}\\
\vspace{1mm} ~~~~~~~~~~~~~~~~~~~~~Fig.3a~~~~~~~~~~~~~~~~~~~~~~~~~~~~~~~~~~~~~~~~~~~~~~~~~~~~~~~~~~~~~~~~~~~~~~Fig.3b~~~\\

\vspace{6mm} Fig. 3a shows the evolution of the deceleration
(red), \emph{jerk} or statefinder (green), \emph{kerk} (brown) and
\emph{lerk} (blue) parameters. The continuous lines represent the
interacting and the dotted lines represent the non-interacting
situations in the emergent scenario.\\
Fig. 3b shows the $r$-$s$ trajectory under interaction and
non-interaction. The trajectories have coincided. We have taken
the cosmic time $t\in[0,4]$. The $r(s)$ trajectory has passed
through the fixed point $\{r=1,s=0\}$ of the $\Lambda$CDM. We have
taken $n=1.1$, $m=2.1$, $a_{0}=0.02$, $\delta=0.05$, $B=3.2$,
$m_{p}^{2}=1$ and $A=2.2$.
\\

\vspace{3mm}

\end{figure}
Using $\rho_{m}$ and $\rho_{G}$ in the first field equation (1) it
is easy to express $H$ as a function $t$ under interaction. Using
this $H$ in the second field equation (2) we get $p_{G}$ under
interaction as

\begin{equation}
\begin{array}{c}
   p_{G}=\left[ -\left(a_{0}\left(B+e^{At}\right)^{m}\right)^{-3(1+w_{m}+\delta)}\rho_{m0}(1+w_{m})-\right.~~~~~~~~~~~~~~~~~~~~~~~~~~~~~~~~~~~~~~~~~~~~~~~~~~~~~~~~~~~~~~~~~~~~~~~~~~~~~~~~~~~~~~~~~~~~~~~~~~~~~~~~~~~~~~~~~~~~~~~~~~~~~~~~~~~~~~~~~~~~~~~~~~~~~~~~~~~~~~~~~~~~~~~~~~~~~~~~~~~~~~~~~~~~~~~~~~~~~~~~~~~~~~~~~~~~~~~~~~~~~~~~~~~~  \\\\
   \frac{3A^{4}a_{0}^{4}\left(1+Be^{-At}\right)^{-4m}\left(B+e^{At}\right)^{4m}m^{4}\xi^{2}}{~_{2}F_{1}[m,m,1+m,-Be^{-At}]^{4}}-\frac{3A^{2}a_{0}^{2}\left(1+Be^{-At}\right)^{-2m}\left(B+e^{At}\right)^{2m}n^{2}m^{2}m_{p}^{2}}{~_{2}F_{1}\left[m,m,1+m,-Be^{-At}\right]^{2}}+\left(2\sqrt{3}\left(1+Be^{At}\right)^{-5m}\right)\times~~~~~~~~~~~~~~~~~~~~~~~~~~~~~~~~~~~~~~~~~~~~~~~~~~~~~~~~~~~~~~~~~~~~~~~~~~~~~~~~~~~~~~~~~~~~~~~~~~~~~~~~~~~~~~~~~~~~~~~~~~~~~~~~~~~~~~~~~~~~~~~~~`\\\\
  \left(B+e^{At}\right)^{-1-3m}\left(a_{0}\left(B+e^{At}\right)^{m}\right)^{-3(w_{m}+\delta)}m\left\{ -4A^{4}a_{0}^{7}\left(B+e^{At}\right)^{1+7m}\left(a_{0}\left(B+e^{At}\right)^{m}\right)^{3(w_{m}+\delta)}\right.\times~~~~~~~~~~~~~~~~~~~~~~~~~~~~~~~~~~~~~~~~~~~~~~~~~~~~~~~~~~~~~~~~~~~~~~~~~~~~~~~~~~~~~~~~~~~~~~~~~~~~~~~~~~~~~~~~~~~~~~~~~~~~~~~~~~~~~~~~~~~~~~~~~~~~~~~~~~~~~~~~~~~~\\\\
  m^{4}\xi^{2}+e^{At}\left(1+Be^{-At}\right)^{5m}(1+w_{m}+\delta)\rho_{m0}~_{2}F_{1}[m,m,1+m,-Be^{-At}]^{5}-~~~~~~~~~~~~~~~~~~~~~~~~~~~~~~~~~~~~~~~~~~~~~~~~~~~~~~~~~~~~~~~~~~~~~~~~~~~~~~~~~~~~~~~~~~~~~~~~~~~~~~~~~~~~~~~~~~~~~~~~~~~~~~~~~~~~~~~~\\\\
  \left.\left. 2A^{2}a_{0}^{5}(1+Be^{-At})^{2m}(B+e^{At})^{1+5m}\left(a_{0}(B+e^{At})^{m}\right)^{3(w_{m}+\delta)}n^{2}m^{2}~_{2}F_{1}[m,m,1+m,-Be^{At}]^{2}m_{p}^{2}\right\}\right]\times~~~~~~~~~~~~~~~~~~~~~~~~~~~~~~~~~~~~~~~~~~~~~~~~~~~~~~~~~~~~~~~~~~~~~~~~~~~~~~~~~~~~~~~~~~~~~~~~~~~~~~~~~~~~~~~~~~~~~~~~~~~~~~~~~~~~~~~~~~~~~~~~~~~~~~~~~~~~~~~~~~~~~~\\\\
  \left(2a_{0}^{3}~_{2}F_{1}[m,m,1+m,-Be^{-At}]^{5}\right)^{-1}\times~~~~~~~~~~~~~~~~~~~~~~~~~~~~~~~~~~~~~~~~~~~~~~~~~~~~~~~~~~~~~~~~~~~~~~~~~~~~~~~~~~~~~~~~~~~~~~~~~~~~~~~~~~~~~~~~~~~~~~~~~~~~~~~~~~~~~~~~~~~~~~~~~~~~~~~~~~~~~~~~~~~~~~~~~~~~~~~~~~~~~~~\\\\
  \left(\left(a_{0}\left(B+e^{At}\right)^{m}\right)^{-3(1+w_{m}+\delta)}+\frac{3A^{4}a_{0}^{4}\left(1+Be^{-At}\right)^{-4m}
\left(B+e^{At}\right)^{4m}m^{4}\xi^{2}}{~_{2}F_{1}\left[m,m,1+m,-Be^{-At}\right]^{4}}+\frac{3A^{2}a_{0}^{2}\left(1+Be^{-At}\right)^{-2m}\left(B+e^{At}\right)^{2m}n^{2}m^{2}m_{p}^{2}}{~_{2}F_{1}\left[m,m,1+m,-Be^{-At}\right]^{2}}\right)^{-1/2}~~~~~~~~~~~~~~~~~~~~~~~~~~~~~~~~~~~~~~~~~~~~~~~~~~~~~~~~~~~~~~~~~~~~~~~~~~~~~~~~~~~~~~~~~~~~~~~~~~~~~~~~~~~~~~~~~~~~~~~~~~~~~~~~~~~~~~~~~~~~~~~~~~~~~~~~~~~~~~~~~~~~~~\\\\
\end{array}
\end{equation}

Using the above forms of $\rho_{G}$, $p_{G}$, $\rho_{m}$ and
$p_{m}=w_{m}\rho_{m}$ we calculate
$w_{total}=\frac{p_{m}+p_{G}}{\rho_{m}+\rho_{G}}$ and plot against
cosmic time $t$ in figure 1. In figure 1 we have considered
$\delta\neq0$ as well as $\delta=0$. Non-zero $\delta$ implies
interaction between dark energy and dark matter, whereas
$\delta=0$ implies the co-existence of dark energy and matter
without interaction. In both of the interacting and
non-interacting situations, it is observed that the equation of
state parameter $w_{total}<-1$, which indicates phantom like
behavior. In figure 2 we have plotted $\rho_{total}$ and
$p_{total}$ against $t$. Here also we have considered both
interacting and non-interacting situation. It is observed that
$p_{total}$ is increasing in the negative direction and
$\rho_{total}$ is increasing with $t$. This indicates that the
energy density is increasing and pressure is decreasing under both
interacting and non-interacting situations.
\\

In figures 3a we have plotted the various parameters
characterizing the accelerating universe against time $t$. We find
that ``lerk" and ``statefinder" parameters are having similar
patterns and are staying at positive level. We also find in this
figure that the deceleration parameter is staying at the negative
level throughout the evolution of the universe. This indicates
that under this interaction in emergent scenario we are getting an
ever accelerating universe. The ``kerk" parameter is also behaving
like the deceleration parameter. The deceleration moving upwards
with evolution of the universe. It may be interpreted from this
pattern that the acceleration is decreasing at late time.
\\

In figure 3b, we have presented the $r(s)$ curve. The trajectory
passes through $\{r=1,s=0\}$ corresponding to $\Lambda$CDM. Also,
we find that the $r$-$s$ trajectory is confined within the first
and fourth quadrant of the $r(s)$ plane and $r$ increases with
decrease in $s$ with the evolution of the universe. The section of
the plot with positive $r$ and $s$ gives the \emph{radiation}
phase of the universe. After passing through the $\Lambda$CDM we
get the end point of the $r(s)$ curve at $r=1.55,s=-0.2$. The
above calculation has been done under the interaction
$(\delta=0.05)$. If we take $\delta=0$, then the trajectory
coincides with that for interacting model. This indicates that the
interaction does not affect the fate of the universe. At this
point this model differs from the interacting new agegraphic dark
energy model proposed by Zhang et al (2010) where the $r(s)$ curve
got the endpoint at $\Lambda$CDM. When we are considering new
agrgraphic dark energy based on generalized uncertainty principle
in the emergent universe we can go beyond $\{r=1,s=0\}$ i.e. the
$\Lambda$CDM.\\\\

\section{Interaction in the \emph{intermediate} scenario}

Interacting dark energy in the intermediate scenario is presented
in this section. For this purpose, using $a(t)=exp (\lambda
t^{\beta})$ in (10) we get

\begin{equation}
\eta=-\frac{\lambda^{-\frac{1}{\beta}}\Gamma\left[\frac{1}{\beta},\lambda
t^{\beta}\right]}{\beta}
\end{equation}

where $\Gamma[x,z]$ is the incomplete gamma function defined by
$\Gamma[x,z]=\int_{z}^{\infty}u^{x-1}e^{-u}du$. Using the above
form of conformal time $\eta$ and using (11), we get

\begin{equation}
\rho_{G}=\frac{3\xi^{2}(\lambda
t^{\beta})^{4/\beta}\beta^{4}}{t^{4}\Gamma\left[\frac{1}{\beta},\lambda
t^{\beta}\right]^{4}}+\frac{3n^{2}(\lambda
t^{\beta})^{2/\beta}\beta^{2}m_{p}^{2}}{t^{2}\Gamma\left[\frac{1}{\beta},\lambda
t^{\beta}\right]^{2}}
\end{equation}

\begin{figure}
\includegraphics[height=2.0in]{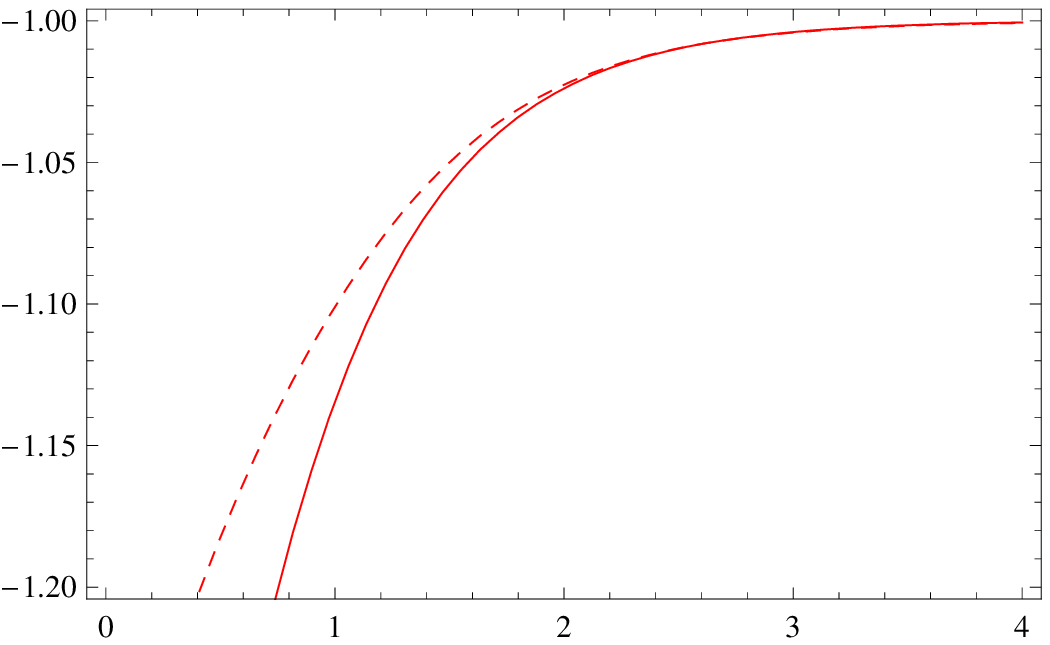}~~~~
\includegraphics[height=2.0in]{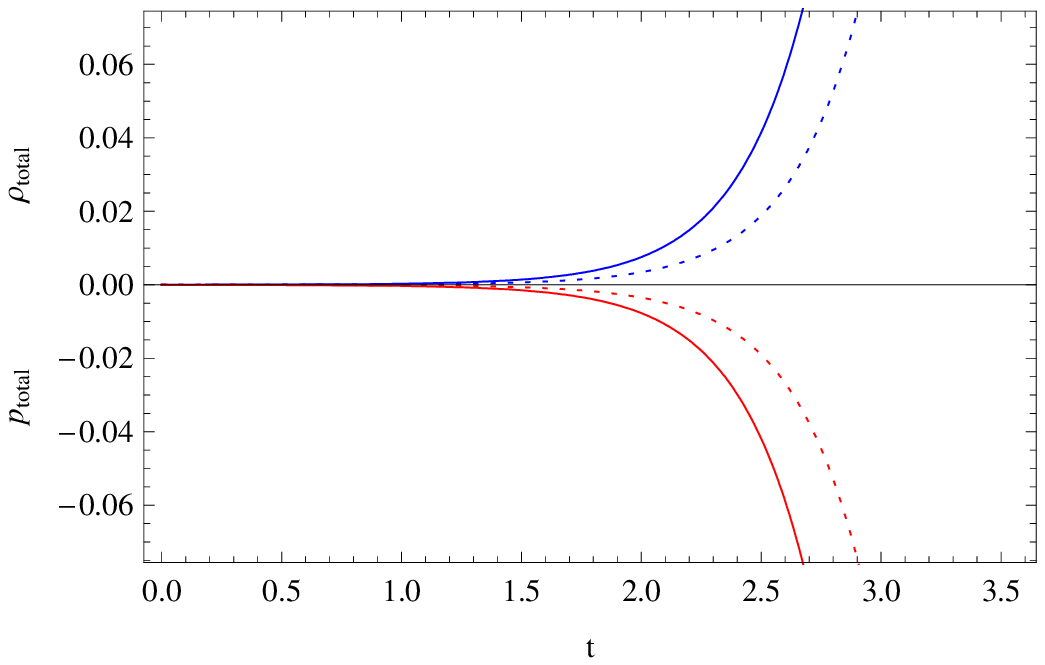}\\
\vspace{1mm} ~~~~~~~Fig.4~~~~~~~~~~~~~~~~~~~~~~~~~~~~~~~~~~~~~~~~~~~~~~~~~~~~~~~~~~~~~~~~~~~~~~~~~Fig.5~~~\\

\vspace{6mm} Fig. 4 shows the evolution of the equation of state
parameter $w_{total}$ in the interacting (thick line) and
non-interacting (dashed line) situations in the
\emph{intermediate}
scenario.\\

Fig. 5 shows the total energy density $\rho_{total}$ and pressure
$p_{total}$ in the interacting (thick line) and non-interacting
(dashed line) situations in the \emph{intermediate} universe scenario.\\

 \vspace{6mm}

 \end{figure}

Using the same technique as in emergent scenario we get the form
of pressure $p_{G}$ under interaction as
 \begin{equation}
 \begin{array}{c}
   p_{G}=\left[-e^{-3\lambda t^{\beta}}(e^{\lambda t^{\beta}})^{-3w_{m}-3\delta}\rho_{m0}(1+w_{m})- \frac{3\xi^{2}(\lambda
t^{\beta})^{4/\beta}\beta^{4}}{t^{4}\Gamma\left[\frac{1}{\beta},\lambda
t^{\beta}\right]^{4}}-\frac{3n^{2}(\lambda
t^{\beta})^{2/\beta}\beta^{2}m_{p}^{2}}{t^{2}\Gamma\left[\frac{1}{\beta},\lambda
t^{\beta}\right]^{2}}\right.\\\\
  -\left\{2\sqrt{3}e^{-3\lambda t^{\beta}}(e^{\lambda t^{\beta}})^{-3w_{m}-3\delta}\beta\left(4e^{2\lambda t^{\beta}}(e^{\lambda t^{\beta}})^{3w_{m}+3\delta}\xi^{2}(\lambda t^{\beta})^{5/\beta}\beta^{4}-\lambda t^{4+\beta}\rho_{m0}\Gamma\left[
   \frac{1}{\beta},\lambda t^{\beta}\right]^{5}(1+w_{m}+\delta)\right.\right.\\\\
 \left.\left.\left. +2e^{\lambda t^{\beta}(2+3w_{m}+3\delta)} n^{2}t^{2}(\lambda t^{\beta})^{3/\beta}\beta^{2}\Gamma\left[\frac{1}{\beta},\lambda t^{\beta}\right]^{2}m_{p}^{2}\right)\right\}\right]\times \\\\
 \left(2t^{5}\Gamma\left[\frac{1}{\beta},\lambda t^{5}\right]^{5}\sqrt{(e^{\lambda t^{\beta}})^{-3(1+w_{m}+\delta)}\rho_{m0}+\frac{3\xi^{2}(\lambda
t^{\beta})^{4/\beta}\beta^{4}}{t^{4}\Gamma\left[\frac{1}{\beta},\lambda
t^{\beta}\right]^{4}}+\frac{3n^{2}(\lambda
t^{\beta})^{2/\beta}\beta^{2}m_{p}^{2}}{t^{2}\Gamma\left[\frac{1}{\beta},\lambda
t^{\beta}\right]^{2}}}\right)^{-1}\\\\
 \end{array}
 \end{equation}

\begin{figure}
\includegraphics[height=2.0in]{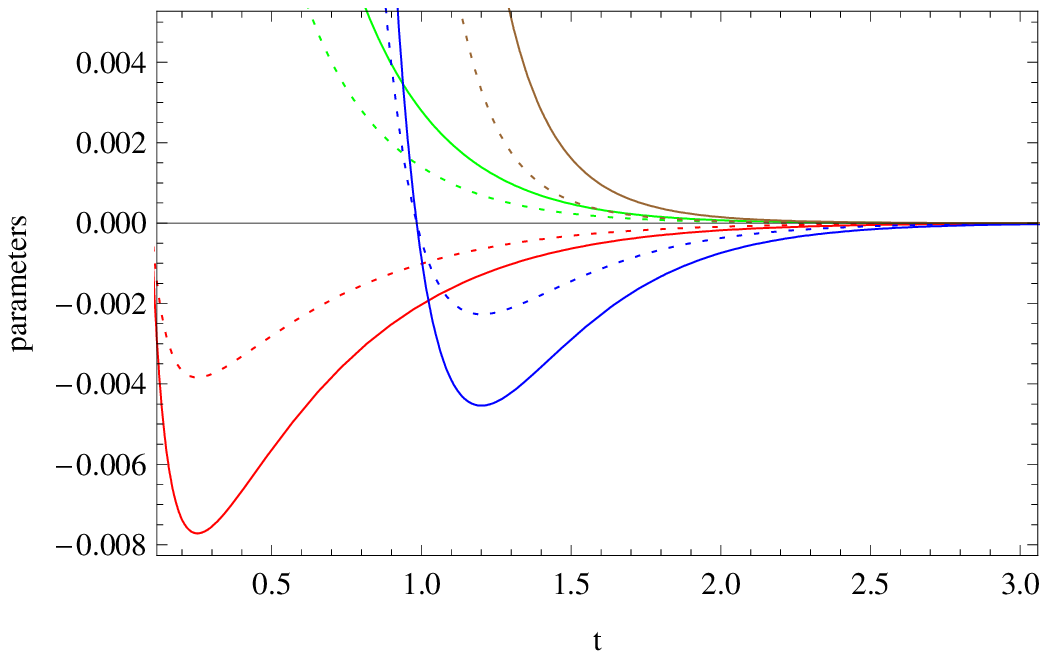}~~~~~
\includegraphics[height=2.0in]{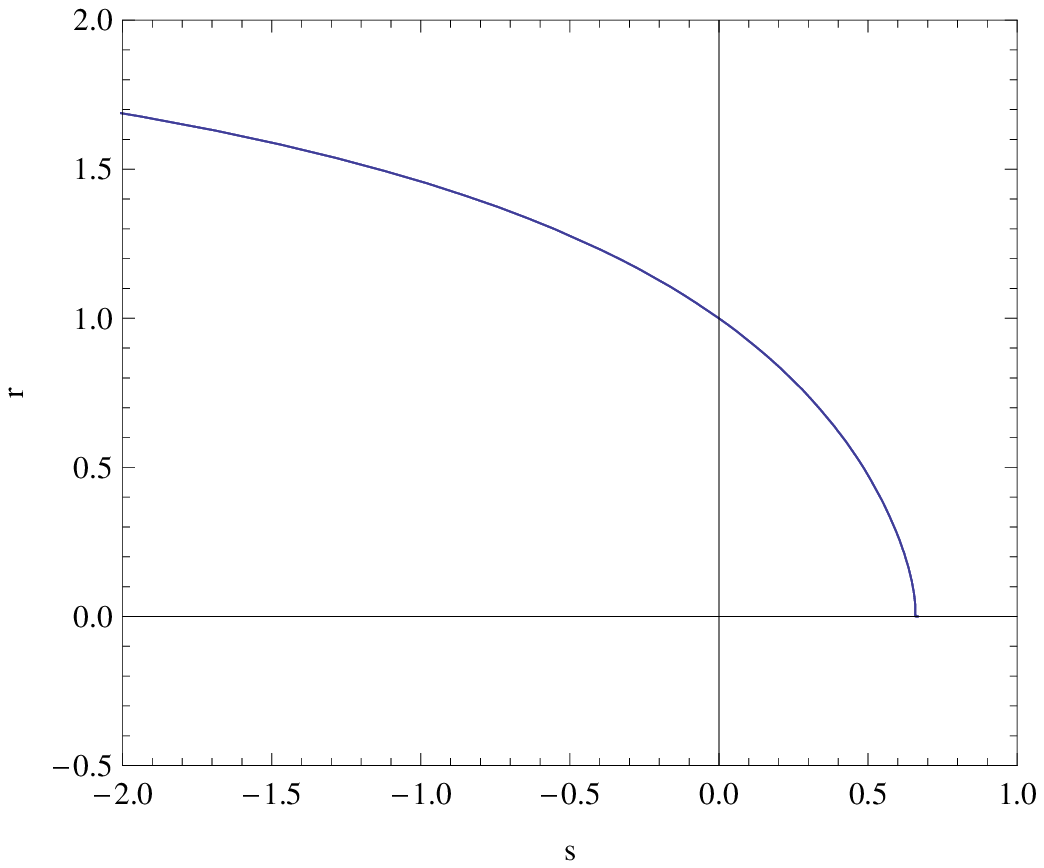}\\
\vspace{1mm} ~~~~~~~~~Fig.6a~~~~~~~~~~~~~~~~~~~~~~~~~~~~~~~~~~~~~~~~~~~~~~~~~~~~~~~~~~~~~~Fig.6b~~~~\\

\vspace{6mm} Fig. 6a shows the evolution of the deceleration
(red), \emph{jerk} or statefinder (green), \emph{kerk} (blue) and
\emph{lerk} (brown) parameters. The thick lines represent the
interacting and the dotted lines represent the non-interacting
situations in the
\emph{intermediate} scenario.\\
Fig. 6b shows the $r$-$s$ trajectory under interaction and
non-interaction. The trajectories have coincided. We have taken
the cosmic time $t\in[0,4]$. The $r(s)$ trajectory has passed
through the fixed point $\{r=1,s=0\}$ of the $\Lambda$CDM. We have
taken $n=2$, $\lambda=1.9$, $\beta=0.5$, $\delta=0.05$ and
$m_{p}^{2}=1$.\\

\vspace{6mm}

\end{figure}

Using the above forms of $p_{G}$ and $\rho_{G}$ we calculate
$w_{total}=\frac{p_{G}+p_{m}}{\rho_{G}+\rho_{m}}$ and plot against
cosmic time $t$ in figure 4. We find that the equation of state
parameter $w_{total}$ is staying below $-1$ in the case of
interaction $(\delta\neq 0)$ as well as non-interaction
$(\delta=0)$. This indicates the phantom-like behavior of the
equation of state parameter. It is tending towards $-1$, but is
not crossing the barrier of $-1$. Hence, it can be said that the
behavior of $w_{total}$ is more or less similar to that in the
case of emergent scenario. In figure 5 we have plotted
$\rho_{total}=\rho_{G}+\rho_{m}$ and $p_{total}=p_{G}+p_{m}$. It
is observed that $\rho_{total}$ is increasing and $p_{total}$ is
increasing in negative direction with time $t$. Here also we get
almost similar behavior to that of the emergent scenario.
\\

In figure 6a we have plotted the various parameters characterizing
the accelerating universe. In this case also we are getting an
ever accelerating universe as suggested by the negative
deceleration parameter. However, there is a difference between its
behavior in the case of emergent and in the \emph{intermediate}
scenario. In the emergent scenario we have seen the deceleration
parameter to increase throughout the evolution of the universe.
However, in the \emph{intermediate} scenario we find it to
decrease in the early epoch and then to increase in the later
stages. This suggests that in this scenario, the acceleration is
increasing fast and then decreasing gradually. This behavior has
been observed in the interacting $(\delta=0.05)$ as well as
non-interacting $(\delta=0)$ situations. The statefinder or jerk
and the lerk parameters are continuously decreasing throughout the
evolution of the universe and are always staying at the positive
level. The kerk parameter transiting from positive to negative
level and then showing asymptotic behavior.\\

The statefinder diagnostics has been presented through $r(s)$
trajectory in figure 6b. Like emergent scenario we are getting the
radiation and the $\Lambda$CDM in this trajectory. However, in the
fourth quadrant of the universe we are seeing that $r$ is finite
with $s\rightarrow-\infty$. This indicates the \emph{dust} phase
of the universe. This could not be derived in the emergent
scenario. In this way, the fate of the universe in the
\emph{intermediate} scenario differs from emergent scenario when
we are considering the new agegraphic dark energy based on
generalized uncertainty principle interacting with dark matter.
The $r(s)$ curve for the non-interacting case has coincided with
the trajectory in the interacting case. Here also, the dark energy
model proposed in the present paper differs from the interacting
new agegraphic dark energy model proposed by Zhang et al (2010)
where the $r(s)$ curve got the endpoint at $\Lambda$CDM. When we
are considering new agrgraphic dark energy based on generalized
uncertainty principle in the \emph{intermediate} scenario we can
get the transition from radiation to $\Lambda$CDM stage through
dust stage.\\

\section{Interaction in the \emph{logamediate} scenario}

In this section we have discussed the interacting dark energy
under consideration in the \emph{logamediate} scenario. For this
purpose, using $a(t)=exp(\mu(\ln t)^{\alpha})$ in (10) we get the
conformal time as

\begin{equation}
\eta=\int\frac{dt}{exp(\mu(\ln
    t)^{\alpha})}
\end{equation}

\begin{figure}
\includegraphics[height=2.0in]{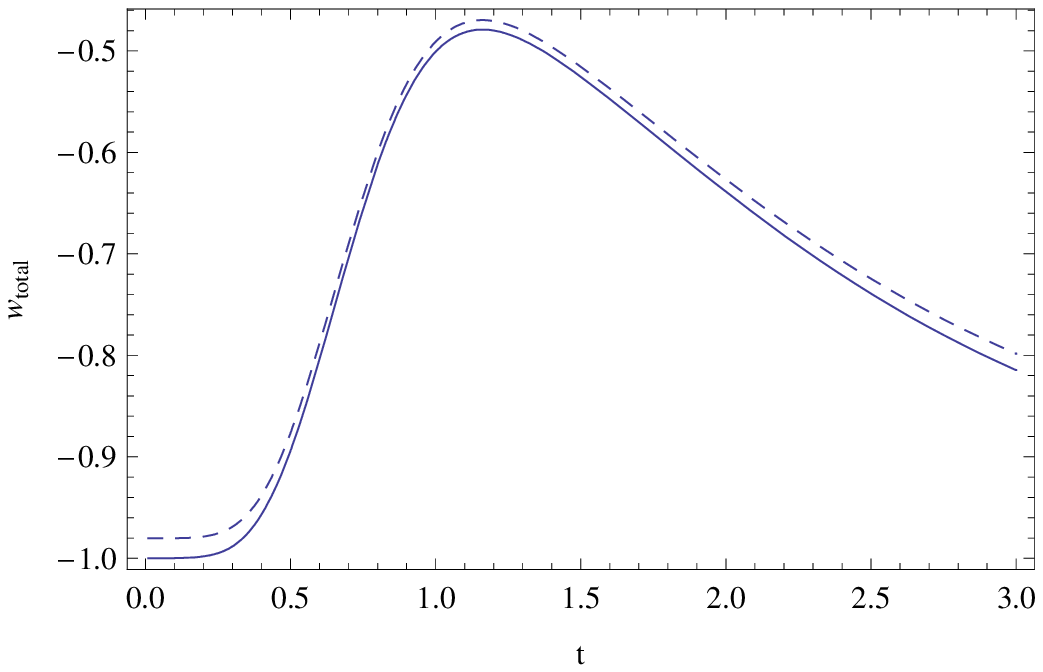}~~~~
\includegraphics[height=2.0in]{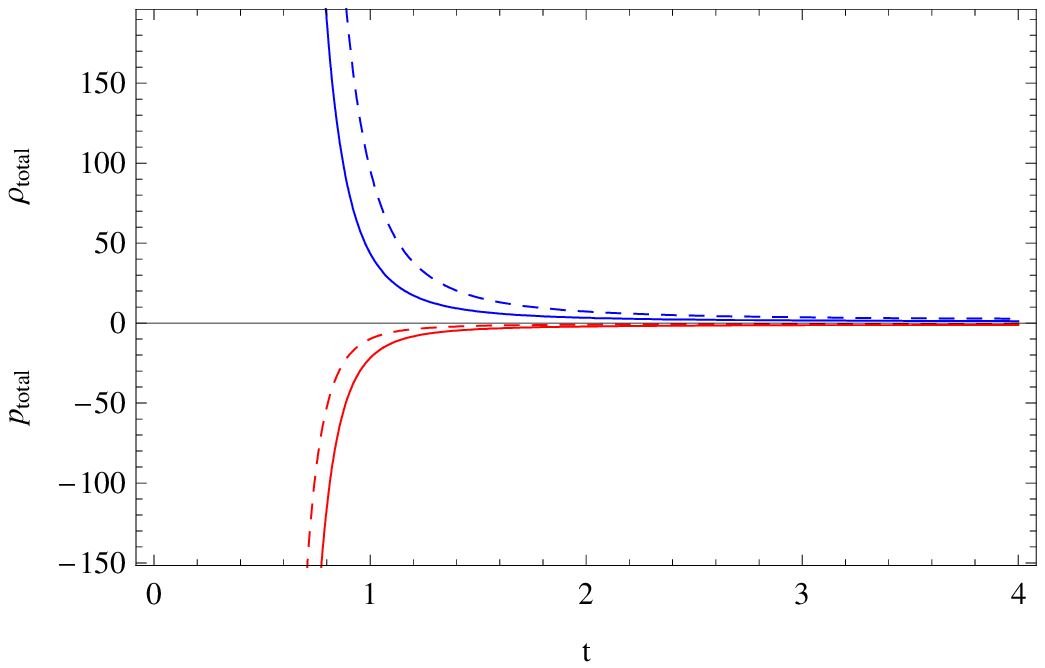}\\
\vspace{1mm} ~~~~~~~Fig.7~~~~~~~~~~~~~~~~~~~~~~~~~~~~~~~~~~~~~~~~~~~~~~~~~~~~~~~~~~~~~~~~~~~~~~~~~Fig.8~~~\\

\vspace{6mm} Fig. 7 shows the evolution of the equation of state
parameter $w_{total}$ in the interacting (thick line) and
non-interacting (dashed line) situations in the \emph{logamediate}
scenario.\\
Fig. 8 shows the total energy density $\rho_{total}$ and pressure
$p_{total}$ in the interacting  (thick line) and non-interacting
(dashed line) situations in the \emph{logamediate} universe scenario.\\

 \vspace{6mm}

 \end{figure}

\begin{figure}
\includegraphics[height=2.0in]{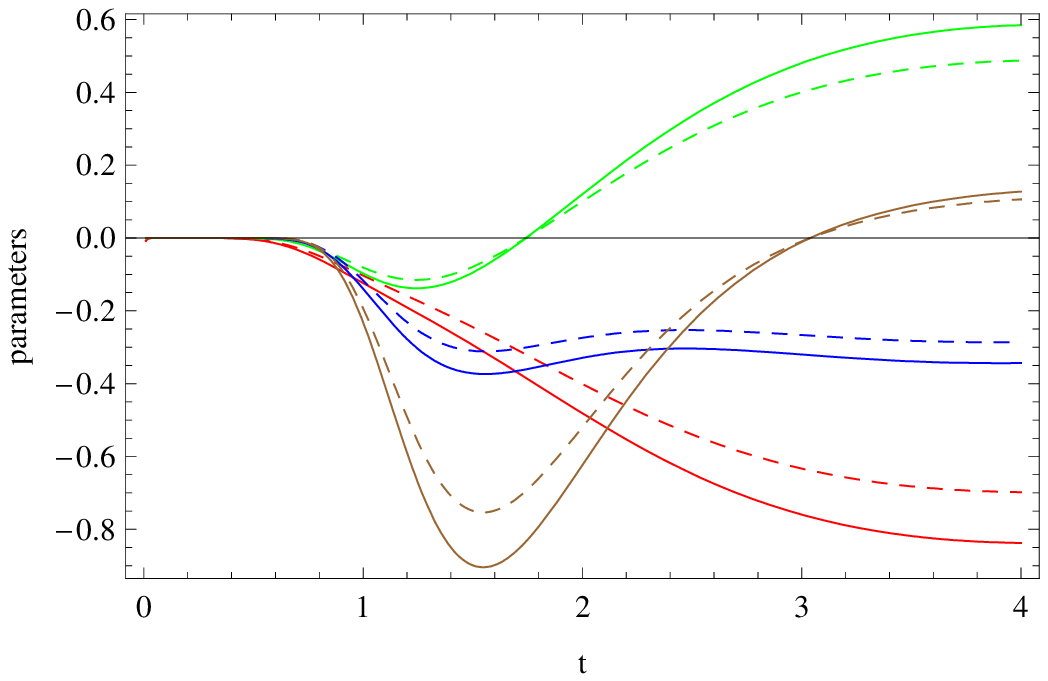}~~~~
\includegraphics[height=2.0in]{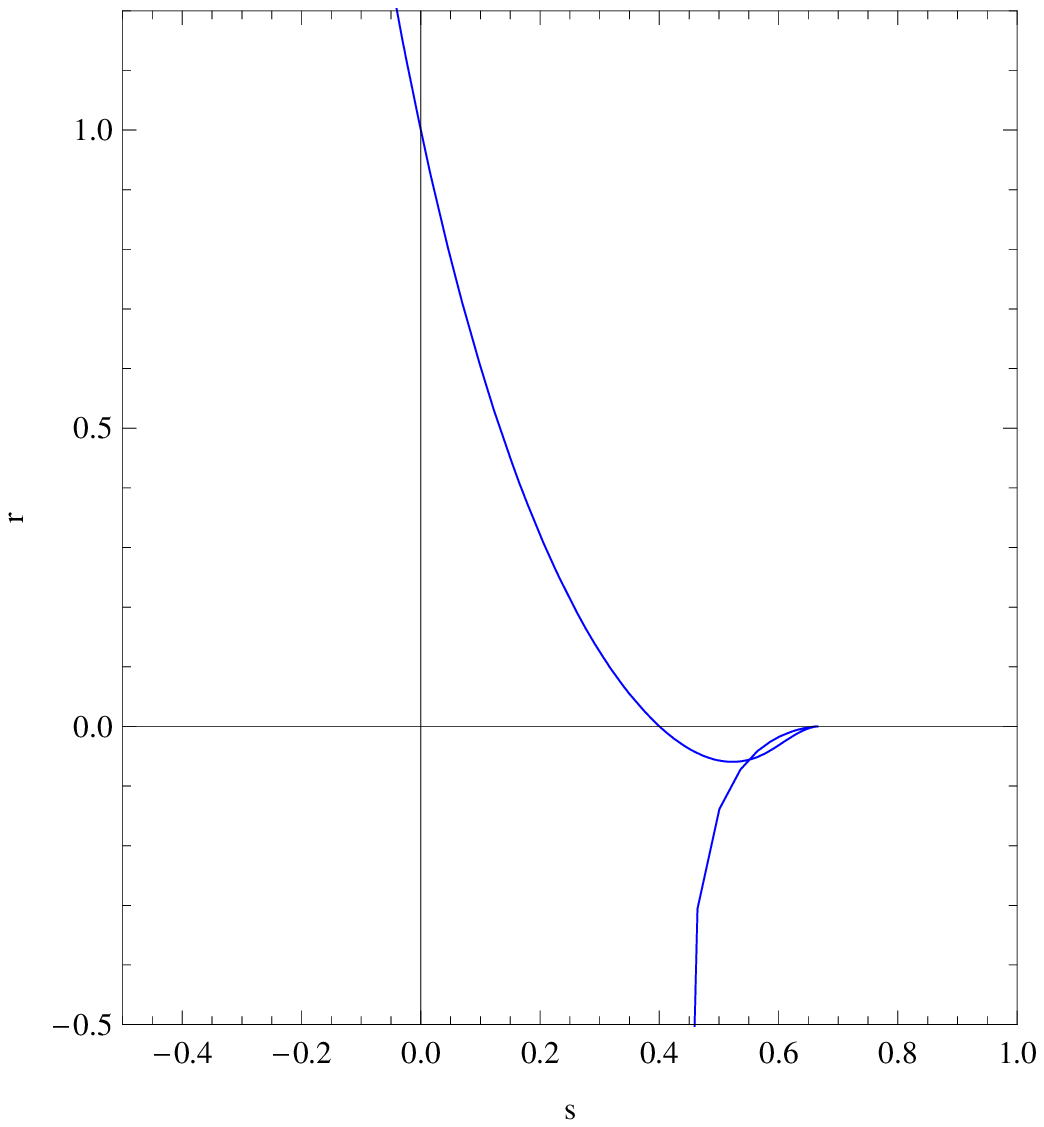}\\
\vspace{1mm} ~~~~~~~~~Fig.9a~~~~~~~~~~~~~~~~~~~~~~~~~~~~~~~~~~~~~~~~~~~~~~~~~~~~~~~~~~~~~~~~Fig.9b~~~~~~~~~~\\

\vspace{6mm} Fig. 9a shows the evolution of the deceleration
(red), \emph{jerk} or statefinder (green), \emph{kerk} (blue) and
\emph{lerk} (brown) parameters. The thick lines represent the
interacting and the dotted lines represent the non-interacting
situations in the
\emph{logamediate} scenario.\\
Fig. 9b shows the $r(s)$ trajectory under interaction and
non-interaction. The trajectories have coincided. We have taken
the cosmic time $t\in[0,4]$. The $r(s)$ trajectory has passed
through the fixed point $\{r=1,s=0\}$ of the $\Lambda$CDM. We have
taken $n=1.0013$, $\delta=0.05$ and $m_{p}^{2}=1$.\\

\vspace{6mm}

\end{figure}

Using this conformal time in (11) and (14) and using (23) we get
$\rho_{G}$ and $p_{G}$. Subsequently we get $\rho_{total}$ and
$p_{total}$ under interaction. Plotting the equation of state
parameter $w_{total}=\frac{p_{total}}{\rho_{total}}$ in figure 7
we view a behavior different from that in the emergent and the
\emph{intermediate} scenarios. We find that $w_{total}\geq -1$
throughout the evolution of the universe. This indicates the
quintessence-like behavior. Whereas, in the emergent and
\emph{intermediate} scenarios we have observed phantom-like
behavior. In figure 8 we find the behaviors of $\rho_{total}$ and
$p_{total}$ deviated significantly from the emergent and
\emph{intermediate} scenarios. Unlike the emergent and
\emph{intermediate} scenarios, the total pressure is decreasing in
the negative direction and energy density is decaying with cosmic
time $t$. Similar behavior has been observed in interacting
$(\delta=0.05)$ and non-interacting $(\delta=0)$ situations.\\

In figure 9a we find that the deceleration parameter is staying at
negative level throughout the evolution of the universe. This
means accelerated expansion of the universe. Moreover the
acceleration is increasing with time. The \emph{jerk} or
statefinder parameter is transiting from negative to positive
sign. The \emph{kerk} parameter is staying at negative level. The
\emph{lerk} parameter is behaving similarly to that of the
\emph{jerk} or statefinder parameter. This parameter is transiting
from negative to positive sign with the evolution of the universe.
Behaviors of the parameters are similar in the cases of
interaction $(\delta=0.05)$ and non-interaction $(\delta=0)$.\\

The statefinder diagnostics has been presented through $r$-$s$
trajectory in figure 9b. Like emergent and \emph{intermediate}
scenarios we are getting the radiation ($r$ and $s$ positive) and
the $\Lambda$CDM i.e. $\{r=1,s=0\}$ in this trajectory. Moreover,
Like emergent scenario, in the second quadrant of the $s-r$ plane
we are seeing that $s$ is finite (negative) with
$r\rightarrow\infty$. There is one characteristic of the $r(s)$
curve that differs from both emergent and \emph{intermediate}
scenarios. At the point $\{s=0.65,r=0\}$ there is a loop that
makes $r\rightarrow-\infty$ with finite $s$ (positive). Here also,
the dark energy model proposed in the present paper differs from
the interacting new agegraphic dark energy model proposed by Zhang
et al (2010).\\\\

\section{Discussions}

In the present work, we have considered the interaction between
the dark energy based on generalized uncertainty principle and
dark matter. We have considered three scenarios, namely, emergent,
intermediate and logamediate scenarios. In figure 1  we have
plotted the equation of state parameter for the interacting dark
energy against cosmic time $t$. This figure shows that the
equation of state parameter $w_{total}$ in both interacting
$(\delta\neq0)$ and non-interacting $(\delta=0)$ situations stays
below $-1$ and then after a certain cosmic time it tends to $-1$
(never crosses $-1$ barrier), which shows phantom like behavior.
Figure 2 shows that the total energy density $\rho_{total}$ is
increasing with time. From the choice of the interaction term it
may be stated that dark matter is getting transferred to dark
energy. Under this situation, the total energy density
$\rho_{total}$ is increasing with evolution of the universe under
emergent scenario. It also shows that the total pressure
$p_{total}$ is increasing in negative direction with cosmic time
$t$ under this scenario. It is apparent from the figure that the
evolution of the $\rho_{total}$ and $p_{total}$ with cosmic time
$t$ in the case of non-interaction are quite similar to that in
the case of interaction.\\

In figure 3a we have plotted the deceleration parameter $q$
against cosmic time $t$ under this interaction. We find that it
stays at negative level in interacting as well as non-interacting
situations. This is consistent with the ever accelerating nature
of the emergent universe. Also we find that $q$ is increasing with
time. This indicates the acceleration of the universe is
decreasing as we move towards late stage of the universe.\\

In figure 4, where we have considered the universe under
\emph{intermediate} scenario, the equation of state parameter (for
$\delta\neq0$) $w_{total}$ is below $-1$ that indicates
phantom-like behavior. However, like emergent scenario it is
tending to $-1$ with evolution of the universe. However, it never
crosses the boundary of $-1$. Similar pattern is available in the
case of non-interaction $(\delta=0)$ situation. From figure 5, we
find the similar behavior of $\rho_{total}$ and $p_{total}$ to
that of emergent scenario. In figure 6a we find that the
deceleration parameter $q$ is negative throughout the evolution of
the universe characterized by \emph{intermediate} scenario.
However, we observe that after a decay up to a certain time, it
starts moving upwards with the evolution of the universe. This
indicates that although the universe is ever accelerating under
\emph{intermediate} scenario, the acceleration itself decreases as
we reach late stage of the universe. However, in the early stage
of universe the
acceleration increased with time.\\

Considering the \emph{logamediate} scenario, we found from figure
7 that the equation of parameter $w_{total}\geq-1$ that indicates
quintessence like behavior. From figure 8 we understood that the
$\rho_{total}$ is decaying and $p_{total}$ is decreasing in the
negative direction. Similar patterns are found in interacting
$(\delta=0.05)$ and non-interacting $(\delta=0)$ situations. Like
the emergent and \emph{intermediate} scenarios the deceleration
parameter $q$ stays at negative level and the acceleration
decreases with time (figure 9a).\\

We have also investigated the statefinder diagnostics in figures
3b, 6b and 9b. We found that the two-component dark energy model
consisting of new agegraphic dark energy based on the generalized
uncertainty principle and dark matter generated $r(s)$ trajectory
passing through $\{r=1,s=0\}$ i.e. $\Lambda$CDM irrespective of
the scenario of the universe and interaction.  In all of the cases
the \emph{radiation} phase is achieved. However, only in the case
of \emph{intermediate} scenario the \emph{dust} phase was
obtained. In all the cases we could go beyond $\Lambda$CDM
contrary to what obtained by Zhang et al (2010) for interacting
new agegraphic dark energy model. Finally, it may be conclude that
the new agegraphic dark energy based on generalized uncertainty
principle coexisting with dark matter behaves like quintessence
era for logamediate expansion and phantom era for emergent and
intermediate expansions
of the universe.\\\\

{\bf Acknowledgement:}\\

Sincere thanks are due to the anonymous reviewers who have given
constructive comments to enhance the quality of the manuscript.
\\\\

{\bf References:}\\
\\
 Arabsalmani, M., Sahni, V.: Phys. Rev. D \textbf{83}, 043501 (2011)\\
 Bachall, N.A., Ostriker, J.P., Perlmutter, S., Steinhardt, P.J.: Science \textbf{284}, 1481 (1999)\\
 Banerjee, N., Das, S.: Gen. Relativ. Grav. \textbf{37}, 1695
 (2005)\\
 Barrow, J. D. and Nunes, N. J.: Phys. Rev. D \textbf{76},043501 (2007)\\
 Barrow, J. D. and  Liddle, A. R.: Phys. Rev. D \textbf{47}, 5219 (1993)\\
 Beck,C., Mackey, M. C.: Int. J. Mod. Phys. D \textbf{17}, 71 (2008)\\
 Bousso, R.: Rev. Mod.  Phys. \textbf{74}, 825 (2002)\\
 Cai, Y-F., Saridakis, E. N., Setare, M. R., Xia, J-Q.: Physics Reports \textbf{493}, 1 (2010)\\
 Campuzano, C., del Campo, S., Herrera, R., Rojas, E., Saavedra,
 J.: Phys. Rev. D \textbf{80}, 123531 (2009)\\
 Calcagni,G. and  Liddle, A. R.: Phys. Rev. D \textbf{74}, 043528 (2006).\\
 Cataldo, M., Mella, P. and  Minning, P., Saavedra, J.: Phys. Lett. B \textbf{662}, 314 (2008)\\
 Chattopadhyay, S., Debnath, U., Chattopadhyay, G.: Astrophys. Space Sci. \textbf{314}, 41 (2008)\\
 Copeland, E.J., Sami, M., Tsujikawa, S.: Int. J. Mod. Phys. D \textbf{15}, 1753 (2006)\\
 Dabrowski, M. P.: Phys. Lett. B \textbf{625}, 184 (2005)\\
 Debnath, U.: Class. Quant. Grav. \textbf{25}, 205019 (2008)\\
 Dunajski, M., Gibbons, G.: Class. Quant. Grav. \textbf{25}, 235012 (2008)\\
 Ellis, G. F. R., Madsen,  M.: Class. Quantum Grav. \textbf{8}, 667
 (1991).\\
 Elizalde, E., Nojiri, S., Odintsov, S. D.: Phys. Rev. D \textbf{70}, 043539
 (2004)\\
 Feinstein, A.: Phys. Rev. D \textbf{66} 063511 (2002)\\
 Feng, C-J., Phys. Lett. B \textbf{670}, 231 (2008)\\
 Garay,L. J.: Int. J. Mod. Phys. A \textbf{10}, 145 (1995)\\
 Gorini,V., Kamenshchik, A., Moschella, U.: Phys. Rev. D \textbf{67}, 063509 (2003)\\
 Guo,Z-K., Piao, Y-S., Zhang, X., Zhang,Y-Z.: Phys. Lett. B \textbf{608}, 177 (2005)\\
 Huang, Z. G.; Song, X. M.; Lu, H. Q.; Fang, W.: Astrophys. Space Sci. \textbf{315} 175 (2008)\\
 Khatua, P. B., Debnath, U.: Astrophys. Space Sci. \textbf{326} 53 (2010).\\
 Kim, Y-W., Lee, H. W., Myung, Y. S., Park, M-I.: Mod. Phys. Lett. A \textbf{23}, 3049 (2008)\\
 Li, M.: Phys. Lett. B \textbf{603}, 1 (2004)\\
 Maziashvili,M.: Phys. Lett. B \textbf{652}, 165 (2007)\\
 Mukherjee, S., Paul, B. C., Dadhich, N. K., Maharaj, S. D., Beesham, A.: Class. Quantum Grav. \textbf{23} 6927 (2006)\\
 Mukherjee, S., Paul, B. C., Dadhich, N. K., Maharaj, S. D., Beesham,
 A.: arXiv:gr-qc/0505103v1(2005)\\
 Nojiri, S., Odintsov,S. D., Tsujikawa, S.: Phys. Rev. D \textbf{71}, 063004 (2005)\\
 Nojiri, S., Odintsov,S. D.: Gen. Relativ. Grav. \textbf{38}, 1285 (2006)\\
 Paul, B. C. and Ghose, S.: Gen. Relativ. Grav. \textbf{42}, 795
 (2010)\\
 Padmanabhan, T.: AIP Conf. Proc. AIP Conf. Proc. \textbf{861}, 179 (2006)\\
 Padmanabhan, T.: Current Science \textbf{88}, 1057 (2005)\\
 Perlmutter, S.J., et al.: Astrophys. J.\textbf{517}, 565 (1999)\\
 Rama, S. K.: Phys. Lett. B \textbf{519}, 103 (2001)\\
 Ratra,B. and Peebles, P. J. E.: Phys. Rev. D \textbf{37}, 3406 (1988)\\
 Sahni,V. Lecture Notes in Physics \textbf{653}, 141 (2005)\\
 Sahni,V. and  Starobinsky, A.: Int. J. Mod. Phys. D \textbf{15}, 2105 (2006)\\
 Sahni, V., Saini, T. D., Starobinsky, A. A., Alam, U.: JETP Lett. \textbf{77}, 201 (2003)\\
 Scardigli, F.: Phys. Lett. B \textbf{452}, 39 (1999)\\
 Seljak, U., Slosar, A., McDonald, P.:  JCAP \textbf{0610}, 014 (2006)\\
 Sheykhi, A.: Phys. Lett. B \textbf{682}, 329 (2010)\\
 Sheykhi, A.: Phys. Lett. B \textbf{680}, 113 (2009)\\
 Visser, M.: Gen. Rel. Grav \textbf{37}, 1541 (2005).\\
 Wang,B., Gong, Y. and  Abdalla, E.: Phys. Lett. B \textbf{624}, 141 (2005)\\
 Wei,H., Cai,R-G., Zeng, D-F.: Class. Quant. Grav. \textbf{22}, 3189 (2005)\\
 Wei, H. and Cai, R-G.: Eur. Phys. J. C \textbf{59}, 99 (2009)\\
 Wei, H. and Cai, R-G.: Phys. Lett. B \textbf{660}, 113 (2008)\\
 Wu, P., Yu, H.: Int. J. Mod. Phys. D \textbf{14}, 1873 (2005)\\
 Zhang, X.: Int. J. Mod. Phys. D \textbf{14}, 1597 (2005)\\
 Zhang, X.: Phys. Lett. B \textbf{611} 1 (2005)\\
 Zhang, L., Cui, J., Zhang, J. and Zhang, X.: Int. J. Mod. Phys.D \textbf{19}, 21 (2010)\\
 Zhao, W.: Int. J. Mod. Phys. D \textbf{17}, 124 (2008)\\

\end{document}